# A Reconfigurable Linear RF Analog Processor for Realizing Microwave Artificial Neural Network

Minning Zhu, *Member, IEEE,* Tzu-Wei Kuo, and Chung-Tse Michael Wu, *Senior Member, IEEE*

*Abstract*—Owing to the data explosion and rapid development of artificial intelligence (AI), particularly deep neural networks (DNNs), the ever-increasing demand for large-scale matrix-vector multiplication has become one of the major issues in machine learning (ML). Training and evaluating such neural networks rely on heavy computational resources, resulting in significant system latency and power consumption. To overcome these issues, analog computing using optical interferometric-based linear processors have recently appeared as promising candidates in accelerating matrix-vector multiplication and lowering power consumption. On the other hand, radio frequency (RF) electromagnetic waves can also exhibit similar advantages as the optical counterpart by performing analog computation at light speed with lower power. Furthermore, RF devices have extra benefits such as lower cost, mature fabrication, and analog-digital mixed design simplicity, which has great potential in realizing affordable, scalable, low latency, low power, near-sensor radio frequency neural network (RFNN) that may greatly enrich RF signal processing capability. In this work, we propose a 2×2 reconfigurable linear RF analog processor in theory and experiment, which can be applied as a matrix multiplier in an artificial neural network (ANN). The proposed device can be utilized to realize a 2×2 simple RFNN for data classification. An 8×8 linear analog processor formed by 28 RFNN devices are also applied in a 4-layer ANN for Modified National Institute of Standards and Technology (MNIST) dataset classification.

## I. INTRODUCTION

DEEP neural network (DNN) as a thriving method in artificial intelligence (AI) and machine learning (ML) has impressively accelerated the development in many applications [1]-[2], including image classification, speech recognition, recommendation system, etc. One of the fundamental building blocks in neural networks is the fully connected neural network, where each neuron **Error! Reference source not found.** is connected to all neurons from the adjacent layer by weighted synapses. A simple multilayer example is shown in Fig. 1, which is inspired by the feedforward architecture of the mammalian visual system [4] as well as other architectures [5]. The information processing in such layered neural structures can be conveniently treated with matrix operations. For each hidden Layer-[*n*] (n>0), the neuron of each layer can be expressed as follows:

$$Z^{[n]}_{q_n \times 1} = W^{[n]}_{q_n \times q_{n-1}} A^{[n-1]}_{q_{n-1} \times 1} + B^{[n]}_{q_n \times 1} \quad (1)$$

$$A^{[n]}_{q_n \times 1} = Activation\{Z^{[n]}_{q_n \times 1}\} \quad (2)$$

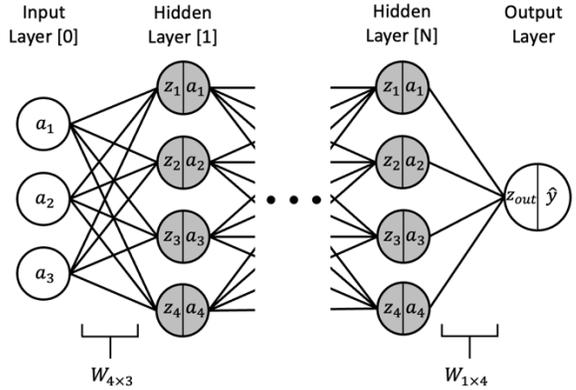

**Fig. 1.** Example of a fully connected neural network.

, where $A^{[n]}_{q_n \times 1}$ in (1) represents the vector of Layer-*n*, which contains $q_n$ neurons and $B^{[n]}_{q_n \times 1}$ is the threshold vector. $W^{[n]}_{q_n \times q_{n-1}}$ represents the weight matrix between Layer-[*n*] and its previous layer, Layer-[*n*-1]. Nonlinear activation functions in (2), such as hyperbolic tangent (*Tanh*), rectified linear unit (*ReLU*), and *Sigmoid* [6] are commonly applied to each hidden layer. As a result, matrix multiplication can be widely applied in both forward (inference) and backward (learning) propagation [1], [7]. Nevertheless, large ML models demand a proliferation of variables. For instance, a powerful neural network that aims to partially function as the human brain can easily scale up to billions of parameters to execute a single task, not to mention the tremendous structure complexity and

Manuscript received April 14, 2023; revised June 21, 2023, and accepted June 30, 2023. This work is sponsored by the Defense Advanced Research Projects Agency under Grant D19AP00030. *(Corresponding author: Chung-Tse Michael Wu.)* Any opinions, findings, and conclusions or recommendations expressed in this material are those of the author(s) and do not necessarily reflect the position or the policy of the Government.

M. Zhu and C.-T. M. Wu are with the Department of Electrical and Computer Engineering, Rutgers University, New Brunswick, NJ 08854 USA (e-mail: minning.zhu@rutgers.edu; ctm.wu@rutgers.edu).
Tzu-Wei Kuo is with the Department of Cell and Developmental Biology, University of Michigan Medical School, Ann Arbor, MI, USA.

connection plasticity of the human brain [8]. Recent analyses show that the demand for computing power will double every 3.4 months, much faster than the rate of Moore's law [9]. Although, in recent years, graphics processor unites (GPU) and the tensor processing unit (TPU) are used to alleviate the computational speed and energy efficiency bottleneck [10], a faster and more efficient realization for matrix multiplication is needed more than ever. On the other hand, conventional von Neumann computing architectures have dominated the modern computer structure design with separated processing and memory units. Nevertheless, it suffers from performing large matrix multiplication when the amount of data shuttling between the processing and memory units increases geometrically, which contributes to at least 50% of the energy dissipation [11].

The spiking neural network (SNN), as a neuromorphic network structure, is one of the non-von Neumann computing structures that combines the processing and the memory units to perform calculations within locality. As such, there is no need to conduct matrix multiplications since all neurons are event driven and can operate locally and asynchronously. Such a structure can harness the two major local-learning features commonly seen in the human brain, the spike-timing-dependent plasticity (STDP) and spike-rate-dependent plasticity (SRDP) [12]. SNN can be realized by both analog and digital approaches. For instance, memristors, discovered by Chua [13] in 1971, can be leveraged to realized analog SNN. By changing the internal ion distribution, memristors can store analog information in the form of resistance [14], which have the potential in realizing neuromorphic computing. On the digital side, CMOS-based technology has been used to design a neuromorphic chip with digital blocks to form a SNN [15]-[17] that emulates the behavior of each neuron in an asynchronous fashion. Due to the complexity of the emulating circuitry, the scalability and power efficiency are limited [18]. Although the SNN model is a progressive approach that aims mimicking the neuromorphic structure of an intelligence being, more fundamental collaborations between ANN designers and biological studies are encouraged [19].

While in-memory computing architecture can allow for acceleration of analog matrix multiplication based on the parallel computing capability, it needs to contain a large number of nonvolatile memory units that can store information precisely in order to perform calculation later, such as memristors or flash memory. It is worth mentioning that memristors can also be used to build high-density crossbar array (CBA), where multiple perpendicular bias lines are used to form a 2D array whose intersection contains a memristor as a memory unit [20]-[22]. In this case, each element of the matrix can be saved and read out by the corresponding two bias lines. Although such structure is very suitable for matrix multiplication, the biasing precision [23] remains one of the major challenges when scaling. On the other hand, while nonvolatile flash memory used in DNNs and SNNs has advantages in mature manufacturing technology, low power, and scalability, it suffers from reliability issues [24].

Recently, the diffraction principle of electromagnetic waves is leveraged to create connections between layers of memorable units in a 3D scenario, which has been applied to analog signal processing form neural networks [25][26]. In particular, this has been achieved by utilizing metasurface arrays, where each layer consists of corresponding spatial modulation [27] inspired by the optical 4f system [28][29]. Although such designs have fast computational speed and power efficiency, they are inevitably bulky due to multiple layers of 2D structures.

In addition, integrated optical designs have recently also been utilized to perform analog complex matrix multiplication with optical networks based on Mach-Zehnder interferometers (MZIs) [30]-[33]. While such a chip-based method is promising in achieving good power efficiency with a small chip size, the fabrication cost and complexity of such an integrated optical device are relatively high. On the other hand, RF/microwave circuits have widely been utilized in analog signal processing [34]. In addition to low power consumption and fast processing speed, RF analog signal processing has many advantages such as low cost, mature fabrication, and ease of analog-digital mixing design. Compared with the optical approach, it requires no additional conversion between electrical and optical signals. While the concept of AI/ML has already been incorporated into RF component design optimization [35] and signal fingerprint enhancement [36], the realization of an RF reconfigurable analog matrix multiplier has not yet been performed to create a microwave artificial neural network or RF neural network (RFNN), which is a promising step in emerging near-sensor and in-sensor computing [37].

To this end, in this work, we propose a reconfigurable quadrature-hybrid-based linear RF analog processor that can perform a set of analog matrix-vector multiplications. The proposed 2×2 reconfigurable linear RF analog processor can be used as part of a multi-layer artificial neural network with additional post data processing for tasks such as data classification and handwriting recognition. The paper is organized as follows: Section II gives a theoretical analysis of the proposed linear RF analog processor. In Section III, a proof-of-concept prototype device is provided, where theoretical analysis and measurement results are introduced. To illustrate how the proposed linear RF analog processor can be utilized in RFNN, in Section IV, the classification capability of a simple $2 \times 2$ RFNN is introduced with theory, simulation, and experimental results. Moreover, the measured S-parameters of the unit cell are utilized to synthesize an 8×8 linear RF analog processor, which is then used in a neural network for handwriting recognition with Modified National Institute of Standards and Technology database (MNIST dataset). Comparison with other approaches as well as possible future research directions for such RF-based neural networks are discussed in Section V.

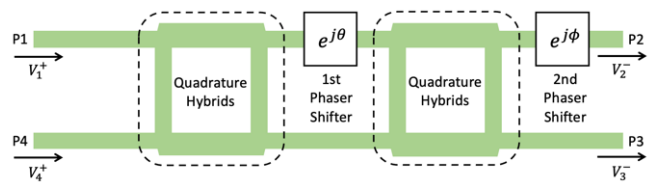

**Fig. 2.** Proposed reconfigurable linear RF analog processor.

## II. METHOD AND ANALYSIS

Analogous to the optical approach based on MZIs, which consists of two beam splitters and two phase shifters [30], the proposed linear RF analog processor consists of two quadrature (90º) hybrids and two phase shifters, as is illustrated in Fig. 2. The quadrature hybrid is a 3-dB (50:50) directional coupler with 90º phase difference between the two output ports. By tuning the first phase shifter, one can split the power from each input port and combine them at the two output ports, correspondingly. Then, the second phase shifter can provide extra phase difference between them. Therefore, the magnitude ratio and phase difference at the output ports can be tuned individually.

The S-parameters of the quadrature hybrid at the design frequency $f_0$ can be expressed as follows [38]:

$$S_{qh} = \frac{-1}{\sqrt{2}} \begin{bmatrix} 0 & j & 1 & 0 \\ j & 0 & 0 & 1 \\ 1 & 0 & 0 & j \\ 0 & 1 & j & 0 \end{bmatrix}. \tag{3}$$

Consider only the voltage components propagating in the forward direction, from ports P1 and P4 to ports P2 and P3, we can express the voltage transformation matrix as:

$$\begin{bmatrix} V_2^- \\ V_3^- \end{bmatrix} = \frac{-1}{\sqrt{2}} \begin{bmatrix} j & 1 \\ 1 & j \end{bmatrix} \begin{bmatrix} V_1^+ \\ V_4^+ \end{bmatrix}. \tag{4}$$

Therefore, the total voltage transformation matrix of the device can be obtained:

$$\begin{bmatrix} V_2^- \\ V_3^- \end{bmatrix} = \begin{bmatrix} e^{-j\phi} & 0 \\ 0 & 1 \end{bmatrix} \frac{-1}{\sqrt{2}} \begin{bmatrix} j & 1 \\ 1 & j \end{bmatrix} \begin{bmatrix} e^{-j\theta} & 0 \\ 0 & 1 \end{bmatrix} \frac{-1}{\sqrt{2}} \begin{bmatrix} j & 1 \\ 1 & j \end{bmatrix} \begin{bmatrix} V_1^+ \\ V_4^+ \end{bmatrix}$$

$$= je^{-j\frac{\theta}{2}} \begin{bmatrix} e^{-j\phi}\sin\left(\frac{\theta}{2}\right) & e^{-j\phi}\cos\left(\frac{\theta}{2}\right) \\ \cos\left(\frac{\theta}{2}\right) & -\sin\left(\frac{\theta}{2}\right) \end{bmatrix} \begin{bmatrix} V_1^+ \\ V_4^+ \end{bmatrix}. \tag{5}$$

One may notice the difference with the optical version of the equation in [30], which is because the phase delay here is defined as a negative value at the output port. The four corresponding S-parameters of the device then can be expressed as:

$$S_{21} = C\, e^{-j\phi} \sin\left(\frac{\theta}{2}\right) \tag{6}$$

$$S_{31} = C \cos\left(\frac{\theta}{2}\right) \tag{7}$$

$$S_{24} = C\, e^{-j\phi} \cos\left(\frac{\theta}{2}\right) \tag{8}$$

$$S_{34} = -C \sin\left(\frac{\theta}{2}\right) \tag{9}$$

, where $C = je^{-j\theta/2}$. When $\theta$ increases from 0 to $\pi$, the device state switches from the cross state (CS) to the bar state (BS), as shown in Fig. 3(a)-(b). For example, assuming $P_1 = 0.5\ mW$, $P_4 = 1.5\ mW$, and both inputs are in phase, the voltage magnitude at port P2 and P3 from port P1 and P4 varies with $\theta$, as shown in Fig. 3(c), which can be expressed as:

$$V_{21} = \sqrt{2Z_0 P_1} S_{21} \tag{10}$$

$$V_{31} = \sqrt{2Z_0 P_1} S_{31} \tag{11}$$

$$V_{24} = \sqrt{2Z_0 P_4} S_{24} \tag{12}$$

$$V_{34} = \sqrt{2Z_0 P_4} S_{34} \tag{13}$$

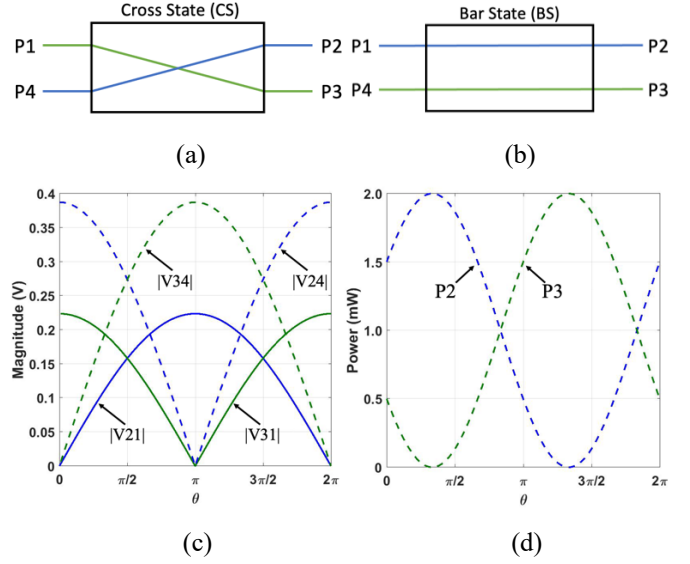

**Fig. 3.** Schematic of two states, (a) cross state, (b) bar state, and corresponding transfer relation of (c) voltage, (d) output power between four ports, when $\theta$ varies from 0 to $2\pi$.

, where $Z_0$ is the characteristic impedance of the transmission line and $V_{nm}$ denotes the voltage magnitude at port $n$ when excited from port $m$. In addition, the power received at ports P2 and P3 can be calculated as follows:

$$P_2 = \frac{|V_{21} + V_{24}|^2}{2Z_0} \tag{14}$$

$$P_3 = \frac{|V_{31} + V_{34}|^2}{2Z_0}. \tag{15}$$

Substituting (10)-(13), we have

$$P_2 = (P_1 + P_4)\sin^2\left(\frac{\theta}{2} + \Delta\right) \tag{16}$$

$$P_3 = (P_1 + P_4)\cos^2\left(\frac{\theta}{2} + \Delta\right) \tag{17}$$

, where $\Delta = \mathrm{acos}\left(\sqrt{P_1}/\sqrt{P_1 + P_4}\right)$. As plotted in Fig. 3(d), the output power, $P_2$ and $P_3$, will also vary with phase $\theta$. When port P2 reaches its maximum power output, the power at port P3 goes to its minimum, and vice versa. Also, from (6) and (8), tuning phase $\phi$ can add extra phase to port P2. Therefore, changing the magnitude ratio and phase difference between port

P2 and P3 can be realized by tuning both phase shifters.

Now, if we consider the linear RF analog processor as a 2×2 transformation matrix $t(\theta, \phi)$, it can be written as:

$$t(\theta, \phi) = \begin{bmatrix} S_{21} & S_{24} \\ S_{31} & S_{34} \end{bmatrix} \text{ and } tt^H = I \quad (18)$$

, which belongs to a particular unitary group of degree two, i.e., $U(2)$ [30]. Since each element of $t(\theta, \phi)$ is not independent of each other, one single device cannot represent an arbitrary matrix. Nevertheless, it can be utilized as the complex unitary matrix in singular value decomposition (SVD) to help synthesize an arbitrary matrix in a cascade fashion, whereas an $N \times N$ matrix can also be formed with multiple such unitary matrices.

## III. EXPERIMENTAL VERIFICATION

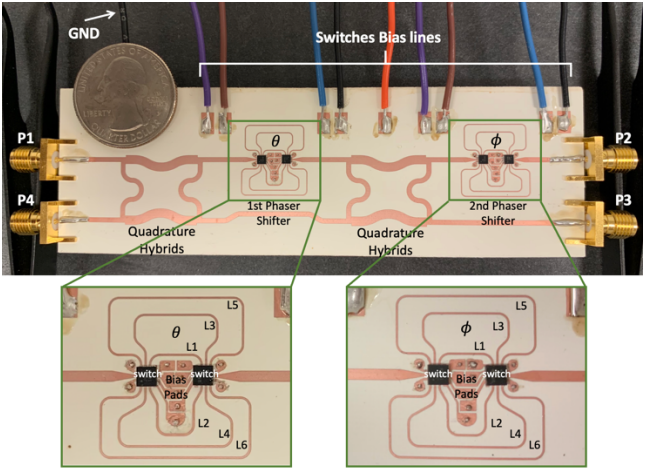

**Fig. 4.** Prototype of the proposed linear RF analog processor (DC biasing lines are routed through vias to the back side of the PCB board).

A prototype of the proposed linear RF analog processor is shown in Fig. 4. The device is fabricated on a Rogers RO4360G2 PCB board with a dielectric constant of 6.15, which consists of two quadrature hybrids with a center frequency at $f_0 = 2$ GHz and two identical discrete phase shifters. Each discrete phase shifter utilizes two SP6T RF switches (Mini-Circuits JSW6-33DR+) that can switch RF signals among 6 different paths with various lengths, labeled as $L_\#$ shown in Fig. 4. The SP6T RF switch contains one input and six outputs, which can be digitally switched using a 3-bit high (~3V) and low (~1.8V) bias voltage signal. Two switches work together to make one of the six transmission lines connected to the rest of the circuit. As such, the device has a total of 36 states since there are 2 phase shifters. Each state can be labeled as $L_n L_m$ and results in a phase difference combination of $\theta_n$ and $\phi_m$, where $\theta_\# = \phi_\# = \beta L_\#$, and $\beta$ is the propagation constant of the transmission line. For the prototype, the discrete phase differences associated with each path are listed in Table I. The switch bias lines are routed and connected through the backside of the board. Fig. 5(a)-(b) show the measured return loss of the device when both phase shifters are at state $L_1 L_1$ or $L_6 L_6$. Fig. 5(c)-(f) indicates how the insertion loss varies with different phase shifter state $L_n L_1$. From $n = 1$ to $6$, $S_{21}$ and $S_{34}$ increases, while $S_{24}$ and $S_{31}$ decreases. Fig. 6 shows how the magnitudes of the insertion loss at 2 GHz varies when tuning the $\theta$ phase shifter. By comparing the values from theory (dashed), simulation (solid), and measurement (plus marks), we can see the signal input from ports P1 and P4 are redistributed to ports P2 and P3, and their magnitude changes with the $\theta$ phase shifter. The simulated and measured data of the prototype also show a similar magnitude shifting tendency. It can be observed that the maximum magnitudes from the simulation and measurement results are lower than the theoretical value, which is due to the loss and phase deviation coming from the imperfect circuit fabrication.

TABLE I
DISCRETE PHASE DIFFERENCE AT $f_0 = 2$GHz

| $\beta L_\#$ | $\beta L_1$ | $\beta L_2$ | $\beta L_3$ | $\beta L_4$ | $\beta L_5$ | $\beta L_6$ |
|---|---|---|---|---|---|---|
| Phase Difference | 29° | 53° | 75° | 104° | 135° | 154° |

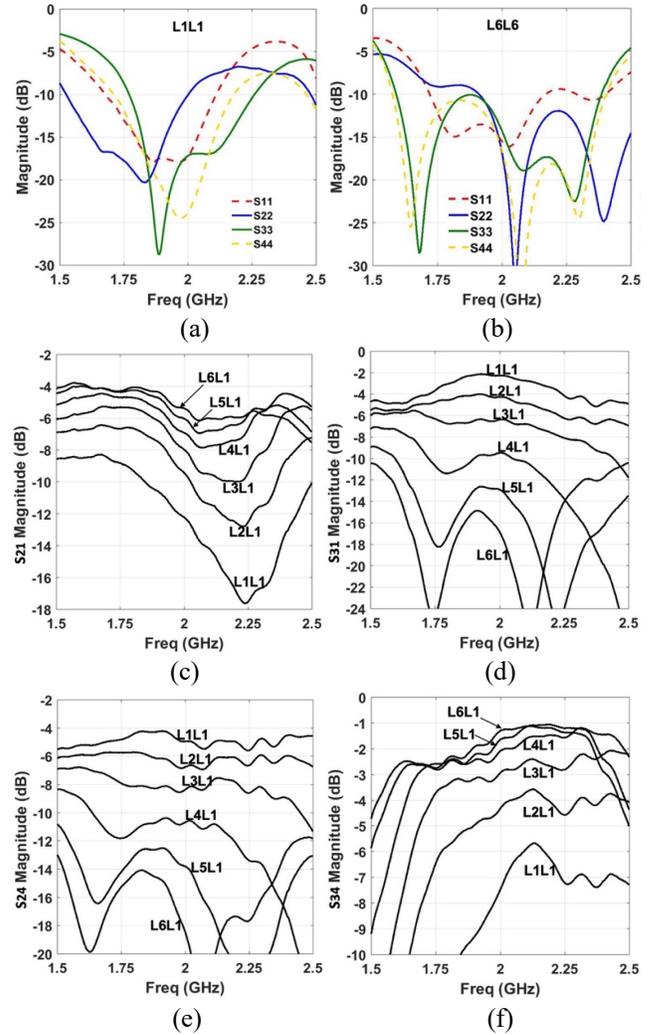

**Fig. 5.** Measured frequency response of the proposed device: return loss of all four ports when the device at (a), state $L_1 L_1$, (b), state $L_6 L_6$, and insertion loss (c) $S_{21}$, (d) $S_{31}$, (e) $S_{24}$, (f) $S_{34}$ when the device at state $L_n L_1$, $n = 1, 2, \dots, 6$.

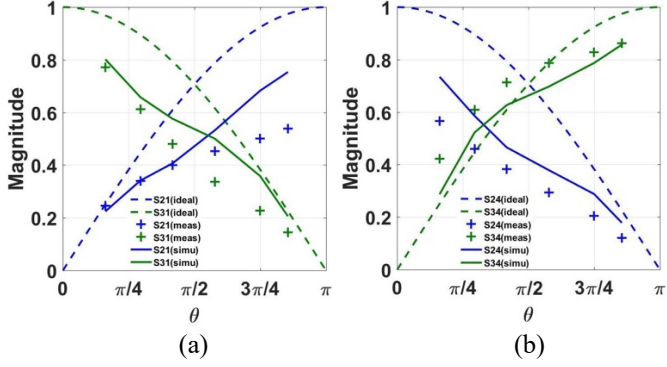

**Fig. 6.** Port P2 (blue) and port P3 (green) S-parameters magnitudes (at 2 GHz) comparison of theory (dashed lines), simulation (solid lines), and measurements ('+'). The $\phi$ phase shifter is at state $L_1$.

## IV. APPLICATIONS

*A. 2×2 Artificial Neural Network*

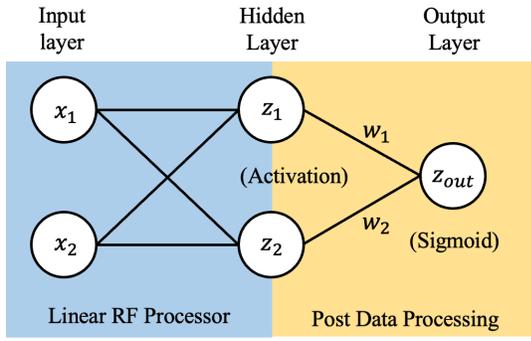

**Fig. 7.** The structure of the 2×2 RFNN.

A 2×2 linear RF analog processor can exhibit good versatility for simple applications. Here, we adopt the matrix as part of a simple 3-layer 2×2 neural network, as illustrated in Fig. 7. The linear RF analog processor shown in Fig. 7 handles the weights multiplication to the hidden layer as follows:

$$\begin{bmatrix} S_{21} & S_{24} \\ S_{31} & S_{34} \end{bmatrix} \times \begin{bmatrix} x_1 \\ x_2 \end{bmatrix} = \begin{bmatrix} Z_1 \\ Z_2 \end{bmatrix} \quad (19)$$

$$[w_1 \quad w_2] \times abs\left\{\begin{bmatrix} Z_1 \\ Z_2 \end{bmatrix}\right\} + b = z_{out} \quad (20)$$

$$Sigmoid(z_{out}) = \hat{y} \quad (21)$$

, where $x_1$ and $x_2$ are inputs, $S_{21}, S_{24}, S_{31}$, and $S_{34}$ are weights between the input and the hidden layer, which can be a positive or negative value. $w_1$ and $w_2$ are weights between the hidden and the output layer, and $b$ is the bias value of the output neuron. Since we measure the magnitude of the device, the absolute function is naturally applied as the nonlinear activation function to the hidden layer as its activation function. As such, Equations (19)-(21) list the entire forward propagation calculation in the neural network illustrated in Fig. 7. The weights between the first and hidden layers are determined by S-parameters of the linear processor, which is reconfigurable with $\theta$ and $\phi$ according to (6)-(9). Once the phase shifter values are determined, one can perform matrix-vector multiplication operation to the weights and the input values, and its result with absolute activation function applied can be measured. It is noted that the operations for the bias in (20), matrix multiplication in (20), and the activation function in (21) for the output layer are conducted in the post data processing. The activation function $Sigmoid(z) = 1/(1 + e^{-z})$ applied to the output layer is commonly used for binary classification [39], such that the final output value $\hat{y}$ ranges between 0 and 1.

For the binary classification purpose, we set the condition as follows: when $\hat{y} \geq 0.5$, it falls to category '1', otherwise, it falls to category '0'. To derive the dividing line, which happens at the critical condition when $\hat{y} = 0.5$ or $z_{out} = 0$, we have:

$$\begin{aligned} z_{out} &= |V_{21} + V_{24}|w_1 \\ &+ |V_{31} + V_{34}|w_2 + b = 0 \end{aligned} \quad (22)$$

since,

$$|V_{21} + V_{24}| = V_1^+ \sin\left(\frac{\theta}{2}\right) + V_4^+ \cos\left(\frac{\theta}{2}\right) \quad (23)$$

$$|V_{31} + V_{34}| = \begin{cases} V_1^+ \cos\left(\frac{\theta}{2}\right) - V_4^+ \sin\left(\frac{\theta}{2}\right) & \text{when } \frac{V_1^+}{V_4^+} \geq tan\frac{\theta}{2} \\ V_4^+ \sin\left(\frac{\theta}{2}\right) - V_1^+ \cos\left(\frac{\theta}{2}\right) & \text{when } \frac{V_1^+}{V_4^+} < tan\frac{\theta}{2} \end{cases} \quad (24)$$

By solving (22)-(24), we obtain two dividing lines separating the classification of '0' and '1', respectively:

$$V_1^+ = \tan\left(\frac{\theta}{2} - \psi\right)V_4^+ + V_L \quad \text{when } \frac{V_1^+}{V_4^+} \geq tan\frac{\theta}{2} \quad (25)$$

$$V_1^+ = \tan\left(\frac{\theta}{2} + \psi\right)V_4^+ + V_S \quad \text{when } \frac{V_1^+}{V_4^+} < tan\frac{\theta}{2} \quad (26)$$

, where

$$V_L = -\frac{b}{w_1 \sin\left(\frac{\theta}{2}\right) + w_2 \cos\left(\frac{\theta}{2}\right)}$$

$$V_S = \frac{b}{w_2 \cos\left(\frac{\theta}{2}\right) - w_1 \sin\left(\frac{\theta}{2}\right)}$$

$$\psi = \text{acos}\left(w_2/\sqrt{w_1^2 + w_2^2}\right).$$

It follows that the 2×2 RFNN can be used for simple binary classification. Fig. 8(a) shows the estimated classification result $\hat{y}$ distribution with the entire input space (The input voltage combinations from ports P1 and P4 range from 0-1V) while applying the absolute activation function to the hidden layer as indicated in (20). In this case, the yellow area represents '1' whereas the blue area represents '0'. The contour lines of values 0.1, 0.5, and 0.9 illustrate a sharp prediction transformation from $\hat{y} = 0$ to $\hat{y} = 1$, which results from the stochastic gradient descent (SGD) optimization for the post-processing parameters [40][41]. Equations (25)-(26) are plotted in Fig. 8(b), which well explains the role of $\theta$ and $\psi$, when determining the shape of the classification region. In particular, $\theta$ determines the orientation of wedge-shape '1' region, while $\psi$ determines the angle of the wedge. Therefore, once given the dataset for binary classification, one can choose the approximate phase value of $\theta$ and half-wedge angle of $\psi$, and calculate for $w_1$, $w_2$, and $b$. This will give a better initial value for further optimization, i.e., the supervised learning process, during which the transformation of the predicted '0' to '1' region will become sharper as the training error keeps decreasing.

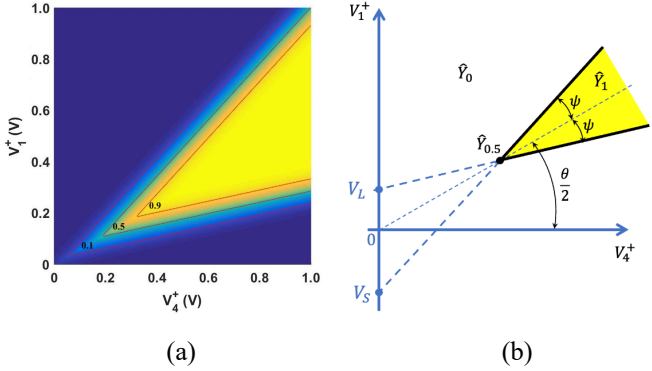

**Fig. 8.** Classification distribution of the neural network (a) and (b) its relation to $\theta, \psi$.

Since the $\phi$ phase does not affect the magnitude at port P2, this leaves us 6 states related to $\theta$. After the learning process, the neural network can automatically operate as six different binary classifiers, in which $\theta$ determines the orientation of the wedge. Fig. 9 shows the $\hat{y}$ of the entire input space, where the x-axis and y-axis are input magnitude $V_4^+$ and $V_1^+$, respectively. The transformation matrix required in (18) is based on the measured S-parameters of the prototype at 2 GHz. Where, the associated states are $L_nL_6$, $n = 1, 2, ..., 6$. For each state, the '0' region (blue area) and '1' region (yellow area) are clearly classified by the 2×2 neural network, i.e., the linear RF analog processor along with post data processing. The contour lines of values 0.1, 0.5, and 0.9 illustrate a sharp evaluation transformation from $\hat{y} = 0$ to $\hat{y} = 1$. Once trained, the RFNN becomes a reconfigurable neural network that can perform 6 different binary classifiers by tuning the $\theta$ phase shifter.

While Fig. 9 is calculated based on the measured S-parameters of the device, to verify the classification capability of the device, we should directly feed in power in port P1 and P4, and then measure the output power at port P2 and P3, to see whether the neural network can classify the data once been trained. For the input space, the x-axis and y-axis are $V_4^+$ and $V_1^+$, respectively. To apply the entire input space of a specific range, we mesh it into 11 by 11 grids, and measure the output power at P2 and P3 with all input power combinations for all six device states, in which the post-processing is conducted on a computer. Once we obtain the optimized parameters, the evaluated $\hat{y}$ distribution of the six different binary classifiers can be plotted as shown in Fig. 10. The patterns at different $\theta$ phase shifter states are similar to those shown in Fig. 9, thereby verifying the classification capability of the proposed RFNN.

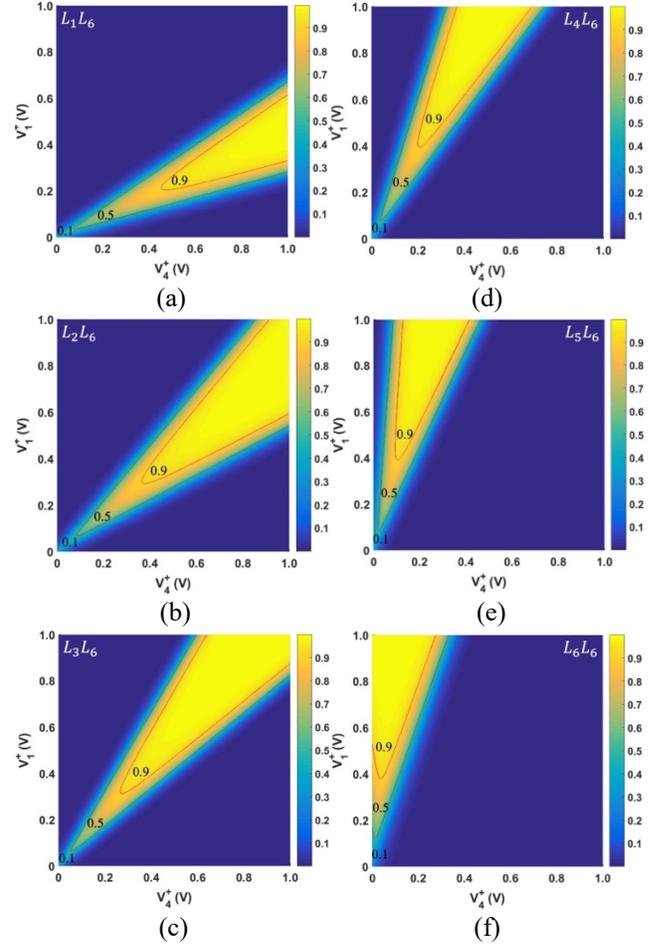

**Fig. 9.** Classification results of the neural network based on the measured S-parameters when the path of the $\theta$ phase shifter switches from $L_1$ to $L_6$.

In addition to the wedge-shape datasets, our simple neural network can also be trained to classify more datasets. Due to the limited number of neurons, the dataset can only be classified with two cuts. The forward/backward propagation during the neural network supervised training process is shown in Fig. 11(a). If necessary, the training dataset $(D_x, D_y)$ can be pre-processed by the computer, including shift and then scale it down with a factor of $\gamma$ to fit the working input space. Then, analog RF power with voltage magnitude equal to $(\gamma D_x, \gamma D_y)$ will be the input signals at P1 and P4. The output power measured in ports P2 and P3 will be converted to voltage magnitude and scaled back by $(\sqrt{2Z_0P_2}/\gamma, \sqrt{2Z_0P_3}/\gamma)$ for further post processing. The predicted value will be used to compare with the ground truth and calculate the error. After one batch of data processed, the accumulated error can be used to update the parameters for the post-processing part, and to update physical parameters on the physical device, such as the biasing voltage, digital biasing code. Such learning process considering real devices can achieve a better performance of a given system [42]. The inference (prediction) process is shown in Fig. 11(b). The trained parameters are configured for the physical device and the post-processing unit. The testing data from the computer will be transformed to input signal and the outcome is measured and post-processed. The prediction by the entire neural network falls into the category of '0' or '1', in which the accuracy can be obtained by comparing the predicted category with the ground truth of the testing dataset.

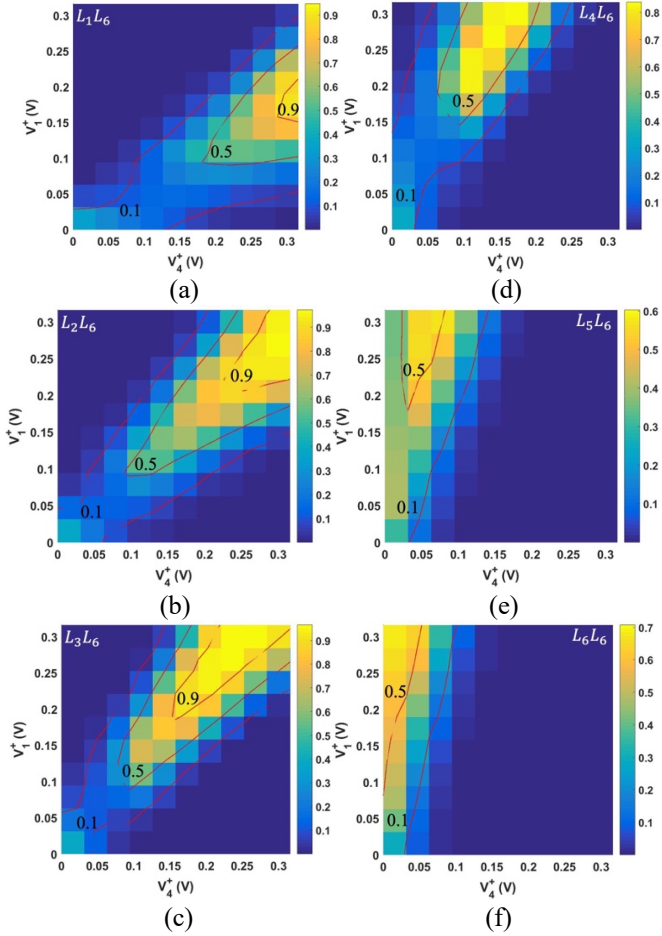

**Fig. 10.** Classification results of the neural network based on the measured output power from port P2 and P3 when the path of the $\theta$ phase shifter switches from $L_1$ to $L_6$.

In addition to the classification results shown in Fig. 10, four more classifiers are trained as illustrative examples, where their classification performance is shown in Fig. 12. The train/test data ranges from 0 to 30, which are multiplied by the scaling factor $\gamma = 1/100$ during the pre-processing and scaled back in the post processing step. It is noted that the data points of label '1' are presented with blue '+' markers, and those of label '0' are with white '*' markers. Fig. 12(a) shows the first case, where the data points labeled with '1' are distributed at the upper right corner and those labeled with '0' are evenly distributed in the rest of space. In this scenario, the neural network picks the state of $L_3L_6$ during the training process (the $\phi$ phase shifter is fixed in all four cases), resulting in an accuracy of around 94% for the test data. In the second case and the third case, shown in Fig. 12(b) and (c), respectively, the data points with different labels are distributed diagonally with a slight overlap, in which one is along the direction to the upper right corner and the other to the lower right corner. In both cases, the classifier provides a good accuracy of 98% and 96%, respectively. For the case in Fig. 12(c), the $\theta$ phase shifter is switched to state $L_4$. In the fourth case shown in Fig. 12(d), the data with label '1' are surrounded by those with label '0', and therefore, it is difficult for a 2×2 neural network to handle the classification. As a result, the accuracy for this case decreases to around 74%. Another reason that can affect the accuracy is that the phase shifter of the prototype can only provide fixed six discrete phase difference, thereby limiting the orientation freedom of the wedge-shape classification region. Therefore, the reconfigurability can be further enhanced with continuous low-loss phase shifters instead of discrete ones. On the other hand, the precision may also be improved when incorporating a larger neural network even with coarse phase resolution, e.g., binary neural network [43].

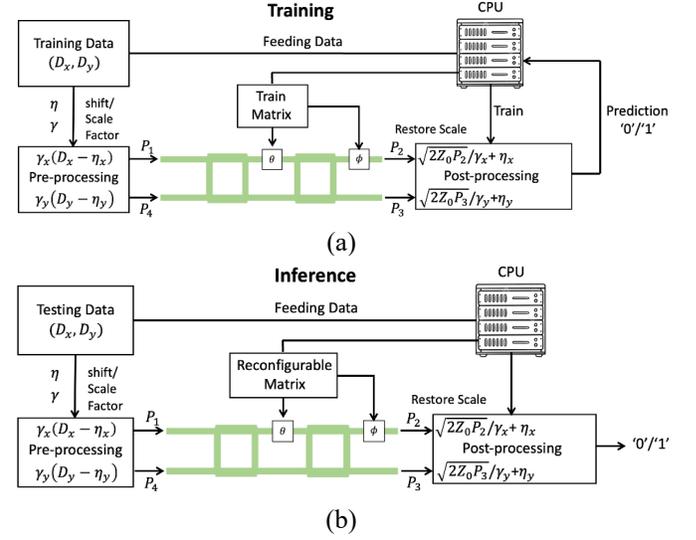

**Fig. 11.** Schematics of the training process and inference (prediction) process of the proposed RF neural network.

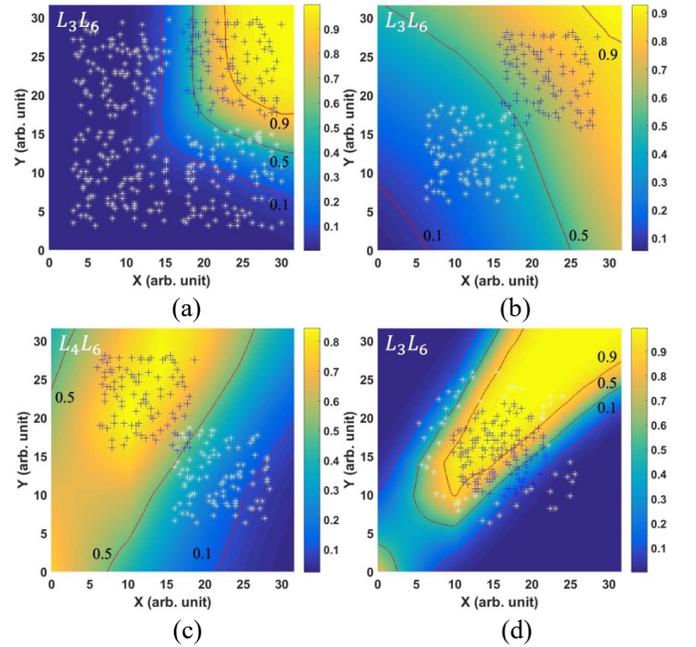

**Fig. 12.** Four examples of the classification results using the proposed RF neural network.

Based on the schematic presented in Fig. 11, it is possible to apply shift, scale, and normalization operations both before and after the data passes through the device. As a result, the device can be treated as a black box that receives data and biasing parameters. Algorithm I outlines the overall training process, wherein the device is optimized using discrete simultaneous perturbation stochastic approximation (DSPSA) [44], while the remaining parameters of the neural network are updated using the stochastic gradient descent (SGD) optimizer.

**Algorithm I** (High-level) Training Process Pseudocode

1: $V \leftarrow$ *Initializing device biasing parameters $V$ randomly*

2: $W \leftarrow$ *Initializing network parameters $W$ randomly*

3: **for** range (**# of epochs**):

4:    **for** range (# of minibatch):   // *minibatch size m*

5:       DSPSA → $\Delta V$; // *calculate $\Delta V$ using DSPSA*

6:       SGD Optimizer → $\Delta W$; // *calculate $\Delta W$ using SGD*

7:       $V = V - \Delta V$; // *update parameters*

8:       $W = W - \alpha \Delta W / m$; // *$\alpha$ is learning rate*

9:    **end**

10: **end**

*B. MNIST Dataset Handwriting Recognition*

The $2 \times 2$ linear RF analog processor can be utilized to synthesize an arbitrary $N \times N$ unitary matrix $U'(N)$, which can be decomposed as:

$$U'(N) = \Sigma(N) \times T(N) \quad (27)$$

, where $\Sigma(N)$ is an $N$-dimensional diagonal matrix with modulus of all its diagonal elements to be 1 and $T(N)$ is also an $N \times N$ unitary matrix that can be further decomposed to an identity matrix by a series of $N$-dimensional rotational matrices $R_{N \times N}^{(i)\,-1}$. If we keep ($N$-2) dimensions unchanged (identical) for each rotational matrix, by multiplying each rotational matrix we are rotating 2 dimensions at a time to a vector in the $N$-dimension Hilbert Space [45][46]:

$$T(N) \cdot R_{N \times N}^{(1)\,-1} \cdot R_{N \times N}^{(2)\,-1} \cdots R_{N \times N}^{(S)\,-1} = I(N) \quad (28)$$

, where $S = N(N-1)/2$. By multiplying $S$ such rotational matrices, we can decompose the entire matrix $T(N)$. Since each rotational matrix is also a unitary matrix, one can easily find its inverse matrix $R_{N \times N}^{(i)} = Hermitian(R_{N \times N}^{(i)\,-1})$ and compose the matrix $T(N)$ back. If each inverse matrix can be realized by analog hardware designs, then one can synthesize arbitrary unitary matrix $U'(N)$, in which $\Sigma(N)$ can be implemented with any devices capable of tuning phase.

The proposed $2 \times 2$ linear RF analog processor can form a unitary matrix and its magnitude and phase at port 2 can be tuned separately, which can be used to synthesize $T(N)$ and $\Sigma(N)$. Fig. 13 shows an example of how to synthesize a $U'(4)$ with the proposed analog processor. The left portion synthesizes the matrix $T(4)$ and the right portion is for the diagonal matrix $\Sigma(N)$. When labeling all processors, one follows the order from left to right, and top to bottom, until all possible columns of processors are filled. Taking the example in Fig. 13, there are five columns of processors in the left portion, each labeled with index $i$ as (1, 2, (3&4), 5, 6). Each processor crosses two adjacent channels leading to corresponding inputs (orange color) and has two tunable phases $\theta_i$ and $\phi_i$ that can be calculated from the required rotational matrix $R_{N \times N}^{(i)\,-1}$ decomposing matrix $T(N)$. Each $N$-dimensional rotational matrix $R_{N \times N}^{(i)\,-1}$ can be formed from an identity matrix replacing four adjacent elements with a unitary matrix along the diagonal, which can be expressed as:

$$R_{N \times N}^{(i)\,-1} = \begin{bmatrix} 1 & \cdots & 0 & 0 & \cdots & 0 \\ \vdots & \ddots & \vdots & \vdots & \ddots & \vdots \\ 0 & \cdots & r_{pp} & r_{pq} & \cdots & 0 \\ 0 & \cdots & r_{qp} & r_{qq} & \cdots & 0 \\ \vdots & \ddots & \vdots & \vdots & \ddots & \vdots \\ 0 & \cdots & 0 & 0 & \cdots & 1 \end{bmatrix} \quad (29)$$

$$\begin{bmatrix} r_{pp} & r_{pq} \\ r_{qp} & r_{qq} \end{bmatrix} = t^H \quad (30)$$

, where $p$ and $q$ are the channel number of the $i$th processor, for example, processor 3 is crossing channel 1 and 2, therefore, $p = 1$ and $q = 2$. Processor 5 is crossing channel 2 and 3, therefore, $p = 2$ and $q = 3$. In practice, the phase value of each processor can be calculated using stochastic optimization methods to avoid imaginary phase condition [30]. Thus, one can easily scale up the design to synthesize an arbitrary $N \times N$ unitary matrix.

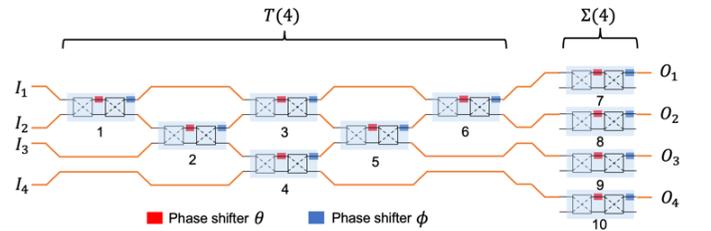

**Fig. 13.** Schematic of a 4×4 linear RF analog processor.

Furthermore, since all real matrices can be decomposed using SVD:

$$M = UDV^H \quad (31)$$

, where $U$, $V$ are unitary matrices, and $D$ is a diagonal matrix, one can synthesize an arbitrary matrix with two unitary matrices as shown in Fig. 13. It is noted that the diagonal matrix can be absorbed into one single matrix in the unitary matrix [47].

The aforementioned approach demonstrates the feasibility of synthesizing an arbitrary matrix using the proposed RF analog processor by carefully selecting device biasing parameters.

9REPLACE THIS LINE WITH YOUR MANUSCRIPT ID NUMBER (DOUBLE-CLICK HERE TO EDIT)However, considering the power loss and limited device biasing space, it is not energy-efficient to synthesize the weighting matrix during the training process. Instead, one can treat the device as a black box or a matrix multiplication accelerator and employ the stochastic method to train the neural network effectively.It follows that by using the stochastic method, a unitary matrix, e.g. $T(N)$ in Fig. 13, contains sufficient parameters to be trained for specific applications. Here, we design an 8×8 matrix $T(8)$ that can be employed in a multilayer fully connected neural network performing classification for handwritten digits in the MNIST dataset as shown in Fig. 14. The weights between hidden Layer-1 and hidden Layer-2 are realized by the 8×8 linear RF analog processor, which is constructed with 28 2×2 RF linear analog processor devices as building blocks shown in Fig. 4. Each 2×2 processor device has 36 different states since there are two phase shifters, and each has 6 discrete phase values.However, considering the power loss and limited device biasing space, it is not energy-efficient to synthesize the weighting matrix during the training process. Instead, one can treat the device as a black box or a matrix multiplication accelerator and employ the stochastic method to train the neural network effectively.

It follows that by using the stochastic method, a unitary matrix, e.g. $T(N)$ in Fig. 13, contains sufficient parameters to be trained for specific applications. Here, we design an 8×8 matrix $T(8)$ that can be employed in a multilayer fully connected neural network performing classification for handwritten digits in the MNIST dataset as shown in Fig. 14. The weights between hidden Layer-1 and hidden Layer-2 are realized by the 8×8 linear RF analog processor, which is constructed with 28 2×2 RF linear analog processor devices as building blocks shown in Fig. 4. Each 2×2 processor device has 36 different states since there are two phase shifters, and each has 6 discrete phase values.

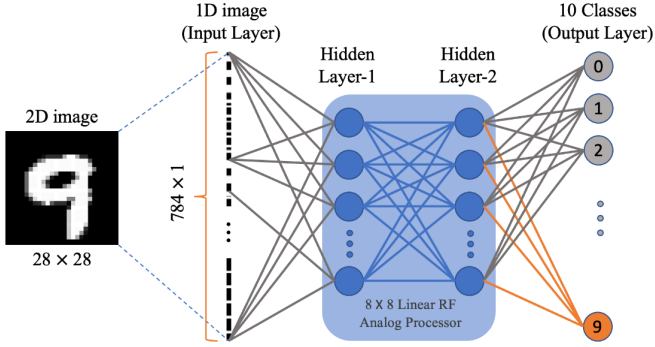

**Fig. 14.** Schematic of the handwriting recognition RFNN.

The entire RFNN has four layers, one input layer, two hidden layers, and one output layer. Each 2D image of 28×28 pixels are reshaped into a 784×1 vector and then compressed into eight features when reaching the first hidden layer. The weights between the fully connected hidden Layer-1 and 2 is synthesized by the 8×8 linear RF analog processor. The output layer maps the eight outputs of the analog processor into 10 classes. From the MNIST dataset, we utilize 50000 images to train the RFNN and the rest 10000 images to test the trained network. The activation functions for the hidden Layer-1 and the output layer are *leaky-ReLU* and *Softmax*, respectively, while the magnitude detection (absolute function) is naturally adopted as the activation function for the hidden Layer-2. In addition, the hidden Layer-2 has no bias parameters. The entire model is built on MATLAB and trained with mini-batch (batch size 10) stochastic gradient descendent method [48]. The training period contains 100 iterations with a learning rate of 0.005 and all training instances are randomly shuffled for each iteration. Although the entire training is done on a computer, the 8×8 linear RF analog processor is simulated based on the measurement data of the unit cell 2×2 linear RF analog processor and constructed based on the same method as shown in the 4×4 case in Fig 13.

The training and testing results are shown in Fig. 15. We compare the RFNN (analog) with the conventional artificial neural network (digital) of the same dimension, where the training accuracy and error in classification of the MNIST training dataset are shown in Fig. 15. It can be observed that the training accuracy of the RFNN is around 91.7%, and the testing accuracy is 91.6%, which is very close to the traditional digital version of the same neural network, with training accuracy of 94.1% and testing accuracy of 93.1%. The performance degradation on the analog version results from the limited discrete phases tunability.

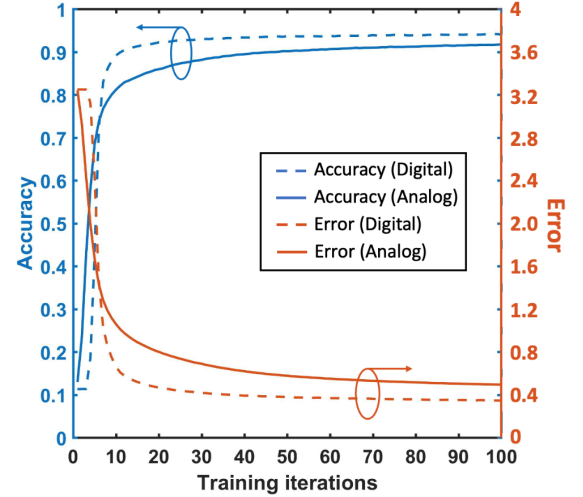

**Fig. 15.** The training accuracy (blue) and error (orange) of the analog (solid line) and digital (dashed line) artificial neural network.

The confusion matrix of the testing dataset for the trained RFNN is shown in Fig. 16. It can be seen that the predicted label and the real label are mostly correct and dominate the diagonal elements of the confusion matrix. The major misclassifications are between the labels of '7' to '9' and '9' to '4', which is classical for the fully connected multi-layer neural networks. Approaches that are specifically designed to handle spatial information and local features, such as convolutional neural networks (CNN), are more suitable for such tasks.

|  | **Predicted Label** | | | | | | | | | |
|---|---|---|---|---|---|---|---|---|---|---|
|  | 0 | 1 | 2 | 3 | 4 | 5 | 6 | 7 | 8 | 9 |
| 0 | 95.36 | 0.0 | 0.4 | 0.61 | 0.1 | 0.81 | 0.81 | 0.4 | 1.31 | 0.2 |
| 1 | 0.0 | 97.09 | 0.56 | 0.56 | 0.0 | 0.38 | 0.19 | 0.19 | 0.94 | 0.09 |
| 2 | 1.31 | 1.51 | 91.83 | 1.31 | 0.3 | 0.1 | 0.61 | 1.01 | 1.61 | 0.3 |
| 3 | 0.97 | 0.49 | 1.26 | 87.96 | 0.0 | 4.66 | 0.0 | 0.39 | 3.3 | 0.97 |
| 4 | 0.41 | 0.92 | 0.1 | 0.0 | 89.93 | 0.1 | 1.22 | 0.51 | 0.81 | 6.0 |
| 5 | 0.98 | 0.77 | 0.77 | 3.28 | 1.64 | 86.89 | 1.86 | 0.44 | 3.06 | 0.33 |
| 6 | 0.41 | 0.52 | 0.52 | 0.0 | 2.07 | 1.55 | 93.69 | 0.0 | 1.24 | 0.0 |
| 7 | 0.37 | 0.73 | 0.83 | 0.83 | 0.18 | 0.28 | 0.0 | 94.68 | 0.09 | 2.02 |
| 8 | 0.3 | 1.49 | 0.69 | 2.18 | 0.0 | 1.78 | 1.09 | 0.1 | 90.39 | 1.98 |
| 9 | 0.83 | 0.31 | 0.21 | 0.73 | 3.54 | 0.52 | 0.0 | 5.31 | 1.35 | 87.2 |

(Real Label on vertical axis)

**Fig. 16.** The confusion matrix for the test dataset, showing the testing accuracy of each label in percentage.

## V. Discussion

Since the EM wavelength in vacuum at the RF region is about $10^5$ larger compared to the infrared light, the physical length of a unit cell linear RF analog processor (Fig. 4) is much larger than that of the unit cell based on MZIs. However, due to the size of two thermos-optical phase modulators and biasing requirement, each optical unit cell is 100$u$m long, which is about 64 times its wavelength (1.545$u$m) [32][49]. On the other hand, the proposed RF unit cell is based on sub-wavelength quadrature hybrids and the length of the unit cell device is roughly one wavelength. It is also worth mentioning that the total size of the RF unit cell can keep decreasing in designs with a smaller center wavelength. Due to the shape of the quadrature hybrids and sufficient router space requirement, one needs to find those low loss substrates that can realize a large transmission line wavelength to width ratio $\chi$, given a fixed 50-Ohm characteristic impedance. Based on the microstrip transmission line theory, such requirement is easier to achieve on a PCB board with thin substrate thickness but high dielectric constant. If we fix $\chi = 100$, with a dielectric constant of 10 and thickness of 0.125mm, we can increase the center frequency easily to $f_0 = 10$GHz. The wavelength in this material is about 12 mm. The typical microstrip transmission line insertion loss on such PCB board is around 0.25-dB per wavelength, which is equal to 5-dB loss per 20 passive 2×2 processor devices in series. A 20×20 passive linear processor array could be very useful for ultrafast signal processing at low cost, since the application of an 8×8 analog matrix formed by 28 unit cells in the handwriting recognition RFNN in previous section has shown its capability. To realize an even larger size of RFNN, one can utilize tensor-train (TT)-decomposed synaptic interconnections to greatly reduce the number of processor devices [50][51] with little precision degradation.

Much lower latency is another benefit for such analog signal processing system, unlike the traditional digital processor, which is restricted to a gigahertz (GHz) clock rate. In addition, the computational complexity of a $N \times N$ matrix-vector multiplication scales as $N^2/p$, which is on the order of $O(N^2)$, where the constant $p$ is the number of very limited parallelisms. However, for analog computing, $p = N$ can be easily realized, and therefore the computation complexity reduces to the order of $O(N)$. The delay can be estimated to be proportional to the length of the processor, since one can roughly assume the signal transmitting in the optical/PCB platform at the speed of light.

When calculating the power consumption for the reconfigurable designs, active components such as phase shifters or RF switches need to be counted in. The RF switch that we use has a power consumption of 0.12-$m$W, which is much smaller than the thermo-optic phase modulator (about 10-$m$W) [32]. The total power consumption synthesizing an $N \times N$ arbitrary unitary matrix, as is shown in Fig. 13, could be as low as $0.12 \times N(N+1) mW$. For specific applications, one may replace the phase shifters with fixed-length passive transmission line once trained and make the entire processor passive. If so, the power consumption of the whole device depends on the sensitivity of the power detector at the output and the power loss of the transmission lines. Since the calculation on such analog devices is much faster than the digital computer, one may compare the power per FLOP (floating points operations) of the RF processor, which is much lower than a digital computer. If we choose the RF power detection rate at the output to be $f_d \approx 10$MHz, which corresponds to $10^7$ $N$-dimensional analog matrix-vector multiplication in one second. This requires a conventional computer to do $2N^2 \times 10^7$ FLOPs per second. Since a typical sensitivity of a RF power detector can be -60dBm, the output power necessary during forward propagation is estimated to be around $10^{-5}Nm$W (consider a 10-dB insertion loss). Therefore, for a passive design, the minimum energy consumption per FLOP of the RF processor scale as $1/(2N)$ $f$J per FLOP, which is much smaller than the convention GPUs (NVIDIA V100) and FPGA (Arria 10), where the power consumption is 31-$p$J per FLOP and 62-$p$J per FLOP, respectively.

TABLE II
COMPARISON AMONG GPU, FPGA, ONN, RFNN, $N = 20$

| Platform | Length (cm) | Unit Cell Length ($\lambda$) | Computation Complexity | Estimated Power Consumption (fJ/FLOP) | Cost | Delay |
|---|---|---|---|---|---|---|
| GPU$ [52] | 30 | NA | $O(N^2)$ | $3.1 \times 10^4$ | Medium | $us$ |
| FPGA& [52] | 24 | NA | $O(N^2)$ | $6.2 \times 10^4$ | Medium | $us$ |
| ONN [32] | 0.76 | 64 | $O(N)$ | 0.25 (passive) | High | $ps$ |
| RFNN# This Work | 46 | 1 | $O(N)$ | 0.025 (passive) | Low | $ns$ |

$ V100;  & Arria 10 FPGA;  # Device estimated with $f_0 = 10\ GHz$

Although in the handwriting recognition neural networks, we utilize four layers of neurons, except for the absolute function in the second hidden layer, we compute all the activation functions in post data processing. In future, by utilizing nonlinear RF device as activation function and power compensation between two linear layers, one may realize multi-layer neural network with such RF processors, which will further enhance its applications. In the optical platform, the optical-to-electrical solution is utilized to realize activation function [53][54]. In the RF circuit platform, one can directly utilize electrical-to-electrical processing solution [55][56]. For example, power detectors and transistors can be used to design non-linear activation function and additional static voltage may serve as bias for each neuron. Such kind of activation also can benefit from the separation of power supply of each layer of neurons. Therefore, it is possible to apply multiple layers of neurons in the neural network. Furthermore, it is possible to realize a submatrix of a given matrix by assigning irrelevant elements as zeros. This capability can be valuable in hyperparameter tuning, including determining the number of neurons and layers in the neural network. Table II lists the merit comparison among RFNN, ONN, FGPA, and traditional digital platform under fixed $N = 20$ and the frequency of the RFNN is assumed to be 10 GHz.

Theoretically, the scalability of such RFNN is subject to the applicable input power and the power detector sensitivity. On the other hand, the number of phase shifters scales with $O(N^2)$, resulting in a rapidly increasing number of biasing lines. This biasing challenge is commonly observed in large-scale reconfigurable planar antenna array systems and can be alleviated using integrated shift-register memory [57],[58]. The design of the RFNN facilitates analog-digital mixed computation, showcasing its potential for a scalable, energy-efficient, low-latency, and learnable RF signal processing capability near or within the sensor, which can be realized on an affordable and mature platform.

VI. CONCLUSION

This work proposes a linear RF analog processor, in which a 2×2 prototype structure is demonstrated for proof-of-concept, which can be used for analog matrix multiplication. The processer has been applied to the hidden layer matrix multiplication of a simple 2×2 neural network with necessary post data processing. Theory analysis, simulation, and measurement results show that the artificial neural network can be successfully trained as a reconfigurable binary classifier with a clear separation of two classes. By tuning the discrete $\theta$ phase shifter, one can apply it to make classifiers oriented in 6 different directions. By varying parameters in post processing, more classifiers can be trained. Moreover, the $\phi$ phase shifter only affects the phase of one output port, which can be used for phase adjusting between the two output ports. The proposed RF analog processer has great potential to be scaled up to realize a low cost, fast, and power-efficient solution for analog matrix multiplication and low latency near-sensor applications. A 4-layer neural network that can do handwriting recognition of the MNIST dataset is also demonstrated. The network utilizes an $8 \times 8$ linear RF analog processor simulated based on the measured S-parameters of the prototype unit cell. Training results show that it can reach test accuracy of 91.6%. Such neural networks working in RF region may also benefit from the mature wireless communication techniques in terms of writing and reading information from each neuron and synapse. Since the 2×2 unit cell device is of sub-wavelength size, one can further miniaturize and integrate the design at higher frequencies. Moreover, machine learning (ML) techniques such as transfer learning and adversarial reprogramming may also be incorporated to enhance their functionalities [59][60]. In addition to fully connected neural networks, other types of neural networks, such as convolutional neural networks (CNN) [61] and recurrent neural networks (RNN) [62], including reservoir computing, can also be implemented using the coherent-based analog processor.